\documentclass[aip,apl,reprint,twocolumn,superscriptaddress,oneside,floatfix,amsmath,showpacs,amssymb]{revtex4-1}

\usepackage{graphicx}
\usepackage{txfonts}
\usepackage{hyperref}
\usepackage{ulem}

\begin{document}


\title{Magnetic field tunable spectral response of kinetic inductance detectors}

\author{F.~Levy-Bertrand}
\email{florence.levy-bertrand@neel.cnrs.fr}
\affiliation{Univ. Grenoble Alpes, CNRS, Grenoble INP, Institut N\'eel, 38000 Grenoble, France}
\affiliation{Groupement d'Int\'er\^et Scientifique KID, Grenoble and Saint Martin d'H\`eres, France}

\author{M.~Calvo}
\affiliation{Univ. Grenoble Alpes, CNRS, Grenoble INP, Institut N\'eel, 38000 Grenoble, France}
\affiliation{Groupement d'Int\'er\^et Scientifique KID, Grenoble and Saint Martin d'H\`eres, France}

\author{U. Chowdhury}
\affiliation{Univ. Grenoble Alpes, CNRS, Grenoble INP, Institut N\'eel, 38000 Grenoble, France}
\affiliation{Groupement d'Int\'er\^et Scientifique KID, Grenoble and Saint Martin d'H\`eres, France}

\author{A.~Gomez}
\affiliation{Centro de Astrobiolog\'ia (CSIC-INTA), Ctra. Torrejon-Ajalvir km.4, 28850 Torrejon de Ardoz, Spain}

\author{J.~Goupy}
\affiliation{CEA/DRF/IRIG – Grenoble  - France}

\author{A.~Monfardini}
\affiliation{Univ. Grenoble Alpes, CNRS, Grenoble INP, Institut N\'eel, 38000 Grenoble, France}
\affiliation{Groupement d'Int\'er\^et Scientifique KID, Grenoble and Saint Martin d'H\`eres, France}

\begin{abstract}
We tune the onset of optical response in aluminium kinetic inductance detectors from a natural cutoff frequency of 90~GHz to 60~GHz by applying an external magnetic field. The change in spectral response is due to the decrease of the superconducting gap, from 90 GHz at zero magnetic field to 60 GHz at a magnetic field of around 3~mT. We characterize the  variation of the superconducting gap, the detector frequency shift and the internal quality factor as a function of the applied field. In principle, the magnetic field tunable response could be used to make spectroscopic measurements. In practice, the internal quality factor behaves hysteretically with the magnetic field due to the presence of vortices in the thin superconducting film. We conclude by discussing possible solutions to achieve spectroscopy measurements using kinetic inductance detectors and magnetic field.
\end{abstract}

\maketitle


Kinetic Inductance Detectors (KID), based on planar superconducting resonators~\cite{Day}, are popular detectors for astrophysical observations~\cite{Concerto,NIKA2,US,Asia,Europe} and interesting devices for physics studies~\cite{axion,BULLKID,sc_coll_modes, Visser, Driessen, Lukas}.
One of the current challenges in millimetre astrophysics observations is to achieve a given degree of spectral resolution without sacrificing the large field-of-view of the current cameras. 
Here we explore a solution to achieve that goal by tuning the spectral response of KID with an external magnetic field. We present the demonstration of the optical response of KID under a variable magnetic field. We also evaluate the effects of the applied magnetic field on the resonators quality factors and conclude by discussing future improvements of our initial KID design.


\begin{figure}
\begin{center}
\includegraphics[width=8.5cm]{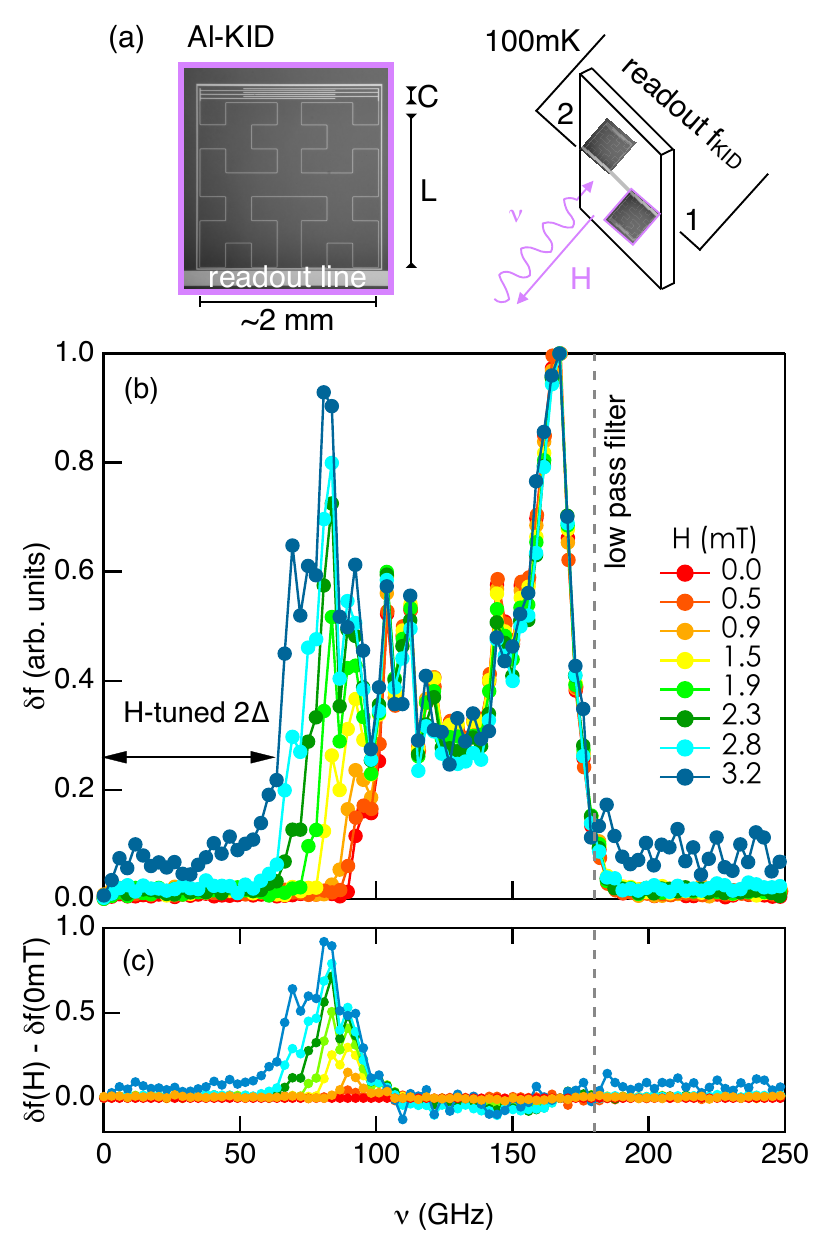}
\caption{\textbf{Magnetic field tunable spectral response of KID.} \textit{(a)} Experimental set-up and picture of a KID. The magnetic field is applied perpendicularly to the array. The illumination is controlled by a Martin-Puplett spectrometer at room temperature and illuminates the array at normal incidence through a series of optical filters and lenses. \textit{(b)} Normalized spectral response of KID to different magnetic fields: frequency shift of  KID as a function the incident optical frequency.  The low cut-off optical frequency corresponds to the $2\Delta$ superconducting gap, which varies as a function of the magnetic field.  \textit{(c)} Difference between the spectral response of KID at different magnetic fields and at zero magnetic field.} 
\label{fig1}
\end{center}
\end{figure}

KID are a particular implementation of superconducting resonators. They are planar $LC$-resonant circuits made of superconductor thin films deposited on an insulating substrate, optimized for photon detection. The photon detection principle consists of monitoring the resonance frequency shift that is proportional to the incident power. Incident radiation breaks Cooper pairs, generating quasi-particles and modifying the kinetic inductance resulting in a shift of the resonance frequency $f=1/(2\pi\sqrt{LC})$ where  $C$ is the capacitance and $L=L_K+L_G$ is the total inductance, i.e. the sum of the kinetic and geometric inductances. The internal quality factor of the resonator also decreases  with the number  of generated quasi-particles.

Figure~\ref{fig1}(a) shows a schematic view of the experimental set-up. The array of KID is made of a 200~nm aluminum film deposited on a high-resistivity silicon wafer. Four wire resistivity measurements performed on the same 200~nm aluminum film give a critical temperature $T_c\sim1.2~$K and a perpendicular critical field $H_c\sim4.5~$mT.  Each KID is coupled via its inductor to the readout line. The inductor $L$ is a Hilbert shape, sensitive to all in-plane polarization~\cite{Monfardini_Hilbert}. The lines are 4~$\mu$m wide. A magnetic field is applied perpendicular to the KID using a custom Helmholtz coil. The KID and the coil are cooled to approximately 100~mK in a dilution refrigerator with optical access. Each coil of the Helmholtz pair has an inner diameter of 50~mm and consists of 7000 turns of NbTi superconducting wire. The magnetic field is quite homogeneous over a volume of 1~cm$^3$ with a gradient of 0.05~mT along the principal axis and 0.025~mT along the radial axis. Calibration measurements with Hall probes gave a current to magnetic field ratio of 18~mT for 100~mA. Illumination is controlled by a Martin-Puplett spectrometer~\cite{MP_suppl} at room temperature and reaches  perpendicularly the KID through a suitable series of optical filters and lenses~\cite{NIKA1}. 


Figure~\ref{fig1}(b) displays the measured spectral response of  KID as a function of the incident optical frequency at different magnetic fields. The response extends from the $2\Delta$ superconducting gap up to the low-pass filter frequency (180~GHz). At zero magnetic field the $2\Delta$ superconducting gap equals 90~GHz, in agreement with the BCS-value  $3.52k_BT_c/h\sim88~$GHz. When increasing the magnetic field the $2\Delta$-gap decreases, increasing the band response from 90-180~GHz to 60-180~GHz at about 3~mT. The level of noise is visible outside the response band, i.e. below $2\Delta$  and above 180~GHz. After normalization of the spectra to have a maximum of 1, the common-band response appears almost identical. Thus, as shown in figure~\ref{fig1}(c), by subtracting the response measured at 3 mT from that measured at zero magnetic field, we can obtain the response of the 60-90~GHz band.  Subtracting the response measured at 0.9~mT from that measured at zero magnetic field shows (already) the possibility of achieving a spectral resolution $R=\nu/\Delta \nu$ of the order of 10. In principle, extrapolating a little further, subtracting the responses measured at very close magnetic fields would give access to the response of a highly resolved spectral band: at 2 mT, a 0.001 mT step, would give a spectral resolution $R=\nu/\Delta \nu$ of about 8000 at 80~ GHz (e.g. a spectral band of 0.01~GHz).

In practice, it is not so straightforward to access the spectroscopic signal. In a magnetic field, the frequency shift is due both to changes of the optical load, the signal of interest, and of the magnetic field. The latter are due to variations in the kinetic inductance $L_k$, resulting from the change of the $2\Delta$ superconducting gap as $L_K=\hbar R_\square/(\pi\Delta)$ where $R_\square$ is the sheet resistance in the normal state~\cite{Annunziata}. 

\begin{figure}
\begin{center}
\includegraphics[width=8.5cm]{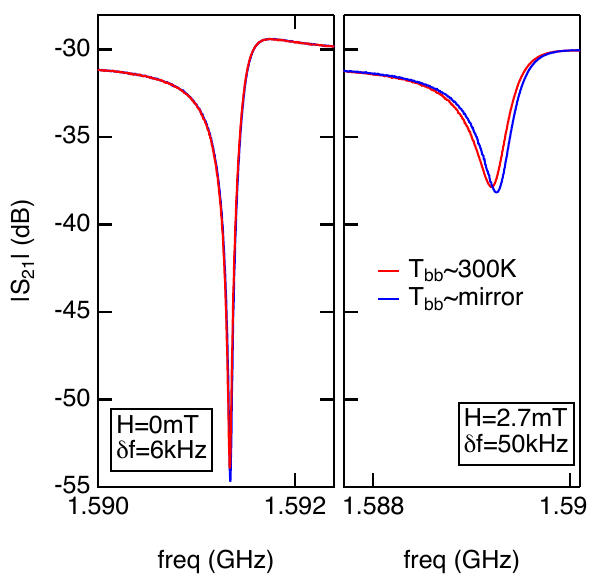}
\caption{\textbf{Magnetic field variation of the VNA response of a KID.} Measurements were carried out under two optical loads, with the cryostat's 300 K window either closed by a dark cap  (emissivity$\sim$1) or closed by a mirror (emissivity$\sim$0). The red curves correspond to the high optical load with the dark cap acting as a blackbody source at about 300~K, $T_{bb}\sim300~$K. The blue curves correspond to the small optical load with the mirror reflecting the emission from the coldest stages of the cryostat, acting as a cold blackbody source, $T_{bb}\sim$~mirror.  \textit{Left} At zero magnetic field, the resonance is deep ($Qi\sim6e4$) and shifts by 6~kHz due to the variation in optical load. \textit{Right} At  almost 3~mT, the resonance is broad ($Qi\sim6e3$) and shifts by 50~kHz. }
\label{fig2}
\end{center}
\end{figure}

Figure~\ref{fig2} illustrates the frequency variations due to both the magnetic field and the change in optical load. The figure shows Vector Network Analyser (VNA) response of a KID under two optical loads and two magnetic fields. 
The red and blue curves correspond, respectively, to measurements under a high or low optical load, with the 300~K window of the cryostat either closed by a dark cap  or closed by a mirror. 
The dark plastic cap acts as blackbody source at about 300~K ($T_{bb}\sim300~$K). 
The mirror, which reflects the emission from the coldest stages of the cryostat, acts as a cold blackbody source ($T_{bb}\sim$~mirror).
Varying the magnetic field from 0 to 2.7~mT, the resonance frequency shifts by about 2~MHz (from left to right panel). Due to the optical signal, the resonance frequency shifts by 6~kHz at 0~mT and by 50~kHz at 2.7~mT.

\begin{figure}
\begin{center}
\includegraphics[width=8.5cm]{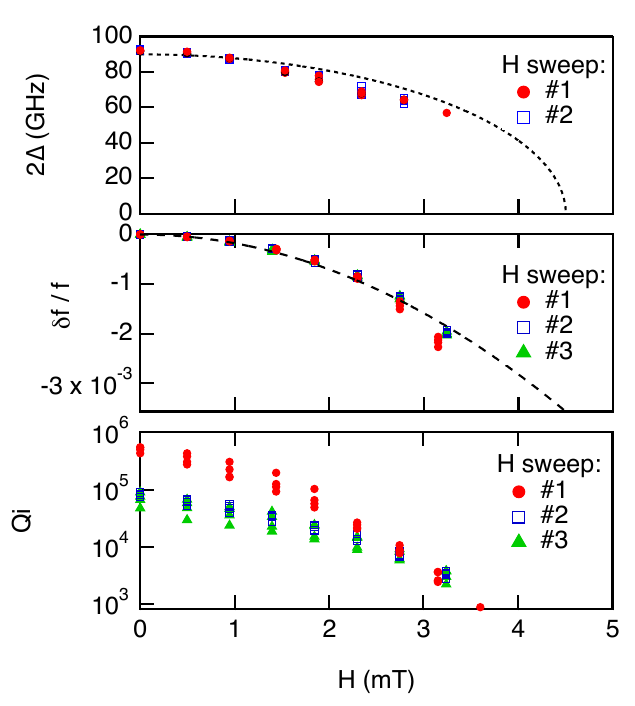}
\caption{\textbf{Magnetic field variation of the superconducting gap, the frequency and the quality factor of KID.} The different points for the same field and field sweep are the response of four different KID.  The change of the $2\Delta$ superconducting gap and of the $\delta f/f$ relative frequency shift with magnetic field  are almost identical for the different magnetic sweeps. For a given magnetic field, the sweep-to-sweep deviation  in $2\Delta$ is less than 1~GHz, and  the deviation in $\delta f/f$ is less than 3e-4. The dash lines are plots of equations (1) and (2). The $Q_i$ internal quality factor depends on the history of the magnetic field.}
\label{fig3}
\end{center}
\end{figure}

To access the absolute spectroscopic signal, preliminary calibrations as a function of magnetic field strength are required. Figure~\ref{fig3} shows the variation of the superconducting gap, the frequency and the quality factor of KID with respect to the magnetic field. The $2\Delta$ superconducting gap and the $\delta f/f$ relative frequency shift remain nearly identical while ramping up and down the magnetic field. The magnetic field dependence of the superconducting gap follows the  formula~\cite{Douglas, Douglas2}:
\begin{eqnarray}
\Delta(H)\sim\Delta_0\sqrt{1-\Bigg(\frac{H}{H_c}\Bigg)^2}
\label{gap}
\end{eqnarray}
where $\Delta_0\sim$~90~GHz is the gap at zero magnetic field and $H_c\sim$~4.5~mT is the critical field. The relative frequency shift under an almost constant small optical load is adjusted with the following formula valid for $\delta f<< f$:
\begin{eqnarray}
\frac{\delta f(H)}{f}\sim-\frac{\alpha_0}{2}\Bigg[1-\sqrt{1-\bigg(\frac{H}{H_c}\bigg)^2}~\Bigg]
\label{deltaf}
\end{eqnarray}
This expression is obtained by combining~\cite{Day} $\delta f/f=-\alpha_0/2\times\delta L_k/L_K$ with the magnetic field dependence of $L_K$. Here $\alpha_0=L_K^0/(L_K^0+L_G)$ is the ratio of the kinetic inductance over the total inductance at zero magnetic field. By fitting the data with equation~\ref{deltaf}, we obtain $\alpha_0\sim$~1.4\%. This value is consistent with the value estimated by radio frequency simulation with the SONNET~\cite{SONNET1, SONNET2} software.  The simulation gives the resonance frequency without kinetic inductance, $f_{L_G}$. The measurement gives the real resonance frequency, $f_{L_K+L_G}$. Values of $\alpha_0\sim1-4\%$ are obtained from the following formula: $\alpha=1-(f_{L_K+L_G}/ f_{L_G})^2$. The value of $\alpha_0$ is low because the kinetic inductance of a 200~nm thick film of aluminum is small~\cite{Adane, Lopez}. Higher kinetic inductance and thus better sensitivity could be obtained with thinner aluminum films. We use relatively thick aluminum here because reducing the thickness from 200 nm to 20 nm would increase the perpendicular critical magnetic field from about 5~mT to 40~mT~\cite{Meservey, phD_Dupre}. For practical reasons, we started with smaller magnetic fields. Future instruments based on the concept of magnetic field tunable spectral response of KID will require thinner superconducting films and higher magnetic fields.

The main current limitation for future instruments is the hysteretic behavior of the internal quality factor, as shown in figure~\ref{fig3}. The internal quality $Q_i$  strongly varies with the magnetic field from about $10^6-10^5$ down to $10^3$ and its value depends on the history of the magnetic field. This behavior, observed in other devices designed for astrophysical applications~\cite{Flanigan, Liu}, is due to the presence of vortices in the resonator meander. The vortices are in the so-called plastic regime, in which they are alternatively pinned and mobile when sweeping the magnetic field~\cite{Borisov, Song}. Vortices develop in aluminum thin films because they are type-II superconductors contrary to bulk aluminum~\cite{Lopez, Tinkham}.

To achieve spectroscopy using kinetic inductance detectors and magnetic field the first point to address is the vortex issue. The theory predicts that in thin superconducting films, vortices develop above a magnetic field perpendicular to the films, $H_0=\pi\Phi_0/(4w^2)$ where $\Phi_0$ is the quantum of magnetic flux and $w$ is the width of the  superconducting line~\cite{Maksimova}. Experiments on 200~nm thick Nb superconducting lines validate the formula~\cite{Stan}, and, show that  the critical field for the vortex nucleation, $H_0$ drops drastically from about 200~mT in bulk Nb to below 1~mT for a 200~nm thick Nb film. For superconducting thin films, the width of the lines must be of the order of 500~nm to avoid the formation of vortices up to 5~mT. Thus, future developments of spectroscopy using kinetic inductance detectors and magnetic field requires a new KID design and e-beam lithography or stepper lithography.

Another point to address is the magnetic field generation. The diameter of the magnetic coil must scale up with the diameter of the field of view which is proportional to the focal plane diameter. One solution is to implement a large standard coil at room temperature, around the cryostat.  The magnetic field at the center of the coil is $H=\mu_0 I N/d$ where $\mu_0$ is the vacuum permittivity, $I$ is the current, $N$ is the number of turns and d is the diameter of the coil. For a diameter of 30~cm, a current of 1~A, and 3000 turns, the magnetic field equals 6~mT. This solution is the simplest from a cryogenic point of view, but the magnetic field on the KID may not be sufficiently homogeneous.  


We tune the spectral response of kinetic inductance detectors with a magnetic field. The change in spectral response is due to the decrease of the superconducting gap when increasing the magnetic field. Our results suggest that it may be possible to achieve high-resolution spectroscopy over a wide field of view using KID and a magnetic field. The main pending limitation is the formation of vortices in the lines of the KID. This can be solved by reducing the line width.  In theory, the spectral resolution could reach several thousands depending also on the superconducting gap steepness. In practice, future spectral resolutions of 100-1000 are likely to be possible. The spectral bands could be adjusted with the superconducting material and the magnetic field. Aluminum films with a critical temperature of $T_c\sim1.2$~K and $H_c\sim5$~mT give access to the relevant 50-100~GHz band. Tantalum with a critical temperature of $T_c\sim4.5$~K and $H_c\sim80$~mT could, in principle, give access to a 200-400~GHz band. The magnetic field tunable response could simply be used to extend the spectral range of the MKID instruments to lower frequencies. Low frequency bands,  i.e. below 100 GHz, may be of particular interest for cosmic microwave background studies, for example to correct for the foreground contributions such as synchrotron emission.

\section*{Acknowledgments}
We acknowledge the contribution of  G. Donnier-Valentin and of T. Gandit, respectively, for the design and for the realization of the Helmotz superconducting coil. We acknowledge the contribution of O. Bourrion for the electronic acquisition. We thank B. Sac\'ep\'e for excellent discussions. We acknowledge the overall support of the Cryogenics and Electronics groups at Institut N\'eel and LPSC. This work has been partially supported by the French National Research Agency through the LabEx FOCUS Grant No. ANR-11-LABX-0013 and the EU\textsc{\char13}s Horizon 2020 research and innovation program under Grant Agreement No. 800923 (SUPERTED).  A. G. acknowledges financial support from PID2022-137779OB-C41 funded by the Spanish MCIN/AEI/10.13039/501100011033.

\section*{Data Availability Statement}
The data that support the findings of this study are available from the corresponding author upon reasonable request.

\section*{REFERENCES}
\bibliographystyle{unsrt}

\end{document}